# Artificial intelligence and the future of diagnostic and therapeutic radiopharmaceutical development: in Silico smart molecular design

**Bahar Ataeinia, MD-MPH† and Pedram Heidari, MD*†**

†Department of Radiology, Massachusetts General Hospital, Boston, MA

**\* Correspondence to:**

Pedram Heidari, MD

**55 Fruit St, Wht 427, Boston, MA 02114**

[Pheidari@mgh.harvard.edu](mailto:Pheidari@mgh.harvard.edu), 617-726-5121

**Key Points:**

- Artificial intelligence (AI) is a novel way for development of new radiopharmaceuticals.
- AI methods enable designing lead radiotracers with favorable pharmacokinetics and pharmacodynamics.
- Using AI methods reduces the radiopharmaceutical development time and cost.
- AI does not obviate the need for in vivo testing of the lead compounds, given that the accuracy of the methods is greatly affected by the assumptions put in the models.
- Radiopharmaceuticals developed by AI have a wide range of applications including central nervous system (CNS), cancer, infection and inflammation imaging.

**Synopsis**

Novel diagnostic and therapeutic radiopharmaceuticals are increasingly becoming a central part of personalized medicine. Continued innovation in the development of new radiopharmaceuticals is key to sustained growth and advancement of precision medicine. Artificial intelligence (AI) has been used in multiple fields of medicine to develop and validate better tools for patient diagnosis and therapy, including in radiopharmaceutical design. In this review, we first discuss common in silico approaches and focus on their utility and challenges in radiopharmaceutical development. Next, we discuss the practical applications of in silico modeling in design of radiopharmaceuticals in various diseases.

**Abbreviation list:**

**Aβ:** Amyloid β

**ADMET:** Absorption, Distribution, Metabolism, Excretion, and Toxicity

**AD:** Alzheimer's Disease

**AI:** Artificial Intelligence

**ANN:** Artificial Neural Networks

**AR:** Adenosine Receptor

**α-syn**: α-synuclein

**BBB:** Blood Brain Barrier

**CAAD:** Computer-Aided Drug Design

**CCK-2R:** Cholecystokinin Receptor Subtype 2

**CNS:** Central Nervous System

**$^{64}$Cu:** $^{64}$Copper

**CXCR4:** Chemokine Receptor-4

**$^{11}$C:** $^{11}$Carbon

**DL:** Deep Learning

**FAK:** Focal Adhesion Kinase

**FBDD:** Fragment-Based Drug Design

**$^{18}$F:** $^{18}$Fluorine

**GGT:** Gamma Glutamyl Transferase

**$^{68}$Ga:** $^{68}$Gallium

**HFPPS:** Human Farnesyl Pyrophosphate Synthase

**HYNIC:** Hydrazinonicotinic acid

**$^{131}$I:** $^{131}$Iodine

**LBDD**: Ligand Based Drug Design

**ML:** Machine Learning

**NMR:** Nuclear Magnetic Resonance

**NOP**: Nociceptin Opioid Peptide

**PD:** Parkinson's Disease

**PET:** Positron Emission Tomography

**PDE:** Phosphodiesterase

**PSMA:** Prostate-Specific Membrane Antigen

**QM:** Quantum Mechanical

**QSAR:** Quantitative Structure-activity Relationships Analysis

**SAR:** Structure-activity Relationships Analysis

**SB:** styrylbenzoxazole

**SBDD:** Structure-Based Drug Design

**SBP:** Structure-based Pharmacophore Modeling

**SSTR2:** Somatostatin Receptor Subtype 2

**SPECT:** Single-photon Emission Computed Tomography

**99m Tc:** $^{99m}$ technetium

**VAP-1:** Vascular Adhesion Protein-1

**2-D:** two Dimensional

**3-D:** Three Dimensional

**Introduction:**

Radiopharmaceuticals play a pivotal role in the rapidly emerging field of personalized medicine. [1] In disease states, altered expression or aggregation of specific molecules allow for non-invasive diagnosis and/or targeted radiotherapy using radiopharmaceuticals. [2-4] Thus, development of new radiopharmaceuticals is key to improved diagnosis and therapy in a broader range of diseases. However, design, validation, and translation of these compounds are time-consuming and require considerable effort and financial investment. [5,6]

A translatable radiopharmaceutical has structural and functional characteristics that result in high affinity for its target, low non-specific binding, and favorable pharmacokinetics. Whether the structure of a compound is novel or similar to an existing compound, substantial effort is required to ensure optimal radiopharmaceutical performance. Despite all the care in designing a compound, there is high likelihood that it fails to adequately engage its target in vivo, due to unpredicted parameters that were not considered during the development phase. [7]

The term "*in silico* medicine" refers to using computer modeling and simulation for conducting biomedical research. In silico approaches can predict outcomes for crucial variables that are quit taxing by conventional in vitro and in vivo radiopharmaceutical development methods. [7,8] Computational models can predict target-binding properties of a compound, and critical pharmacokinetic characteristics such as absorption, distribution, metabolism, excretion, and toxicity (ADMET). [9] Therefore, incorporation of in silico methods could facilitate faster and more cost-effective design and testing of new radiopharmaceuticals, by informing in vitro and in vivo studies which reduces the need for animal models to evaluate the lead compounds. [10-14]

Artificial intelligence (AI) methods have been used to improve drug discovery since the 90s. [15] With ever growing rapid expansion of our knowledge about the structure and function of biological targets, AI can expedite radiopharmaceutical design research resulting in more rapid incorporation of these compounds in routine medical practice (Figure 1). [16-18] It is important to note that computational modeling is not a replacement for in lab experiments but rather a complementary assist tool to facilitate radiopharmaceutical development as currently no modeling can emulate the complexity of human body.[12,19]

Herein, we summarize the current computer-aided drug design (CADD) methods, their potential application, challenges in radiopharmaceutical development research and provide real-world examples of radiopharmaceutical design using the input from CADD methods.

**Approach**

In general, in silico approaches focus on structure and behavior of compounds for modeling (Figure 2). Structural models predict binding affinity of a radiopharmaceutical for its target, while behavioral approaches focus on its pharmacokinetics. Both approaches are complimentary and essential since a translatable radiopharmaceutical should have high specificity and affinity for the target, optimal biodistribution, stability, practical effective half-life and a desirable clearance kinetic based on its intended diagnostic or therapeutic applications.[20]

Structure-based computational modeling approaches are classified into two main categories, structure-based drug design (SBDD) and ligand-based drug design (LBDD). SBDD is used to predict radiopharmaceutical-target interaction when the structure of the target is known. For instance, chemical structure of some of the most studied targets such as prostate-specific membrane antigen (PSMA) and chemokine receptor-4 (CXCR4) is known and can be used for modeling in SBDD. [21,22] If the structure of the target is not available LBDD models can be used, which analyzes the structure of known ligands for the target (Figure 3). [12,23,24] Here we discuss structural and behavioral approaches in more detail.

**Structural computational modeling**

1. *Structure-based drug design (SBDD) (direct approach)*

    1.1. *Molecular docking*

    Molecular docking models predict candidate ligands' interaction to a target with a known three-dimensional (3-D) structure. The model predicts various orientations and conformations of ligand and target to determine the ideal pose with minimum energy for best-fit interaction and complex stability. Moreover, each ligand's binding affinity can be computed based on binding energetics by the algorithm's scoring functions to rank the ligands. [25,26] Scoring functions are categorized as classical, and machine learning (ML) functions. [27,28] Determination

of the appropriate approach depends on the target and data availability. [25,29] Different AI techniques such as swarm intelligence, ant colony optimization algorithms, and DL have been used for molecular docking.[30-32] Selecting the most appropriate software and scoring function is a critical step and can significantly affect outcomes.[33] Docking models cannot be reliably used if chemical structure of the ligand or the specific binding site are unknown, unless protein homology modeling and binding site identification is performed prior to docking. [34-36] Molecular docking models are commonly used in radiopharmaceutical development studies to predict the effect of labeling with different radionuclides and chelators on ligand's binding affinity to the target. [37-41]

### 1.2. Fragment-based drug design (FBDD)

This method is an alternative for the modeling of compounds with high molecular weight and low solubility. In FBDD, low molecular weight fragments that interact with biological targets are first identified, and appropriate fragments are then expanded by adding chemical groups or merging with other fragments to design the final compound. [42,43]

### 1.3. Pharmacophore modeling

Pharmacophore analysis is applicable in both SBDD and LBDD, although is more commonly used with the latter. This method mainly focuses on the molecular features, important for the target-ligand interaction. Such features include hydrophobic groups, aromatic rings, polar groups, and hydrogen bond vectors. [23,44] Structure-based pharmacophore modeling (SBP) is a SBDD pharmacophore analysis which is used if the 3-D structure of the target is known. The model can include structural properties of both target and ligand or the target exclusively. [45,46] LBDD pharmacophore modeling is discussed in more details in section 2.2.

## 2. ligand-based drug design (LBDD) (indirect approach)

### 2.1. Ligand chemical similarity search

In this method, two-dimensional (2-D) molecular fingerprints of an already known ligand are used to identify candidate compounds from a database of chemicals with a similar structure to a known ligand. Molecular fingerprints are binary codes that record the presence or lack of specific chemical groups and can be utilized by screening software to perform the chemical similarity search. [23,47] Given that a compound's complete structure is used in chemical similarity search in contrast to other LBDD methods, such as pharmacophore analysis, it may provide a more accurate estimation of in vivo biological activity.

### 2.2. Pharmacophore analysis

As previously mentioned, this approach incorporates crucial features for ligand-target interaction into the model. Initially, a wide range of biologically active ligands is studied to train the model. The main two goals are to distinguish overlapping structures for optimal ligand-target interaction and create conformational space for the studied ligands to determine structural flexibility in the model. However, there are several challenges in designing a pharmacophore model including the modeling of ligand flexibility, molecular alignment and proper selection of training set compounds, which is even more challenging in designing radiopharmaceuticals.[23,44] Even when a radionuclide is added to the backbone structure and not the active binding site, it can alter the radiopharmaceutical's biological behavior. [48-53] Pharmacophore analyses is mainly utilized to screen potential ligands or to design a new radiopharmaceutical. [44] Pharmacophore models can also be used in SBDD approaches as previously discussed.

2.3. *Structure-activity relationships analysis (SAR)*

SAR is based on the assumption that molecules with similar chemical structures are more likely to demonstrate a similar biological activity. [54,55] Using a regression model that correlates structural properties of a known group of similar ligands to their biological activity, such as binding to the desired target or inhibiting an enzyme, SAR can predict a pharmaceutical's biological behavior. [56] Quantitative structure-activity relationship (QSAR) is a SAR that quantifies bioactivity level of a pharmaceutical. Using artificial neural networks (ANN) and deep learning (DL) algorithms, QSAR has been used for drug design. [15,57] The crucial step in SAR is carefully choosing molecular properties used in the regression model to avoid overfitting. An overfitted model has optimal performance on the training dataset but fails when used on new data. [15] Note should be made that structural similarity does not necessarily imply similar activity indicating the need for cautious use of this assumption given the complexity of in vivo biochemical interactions. [58] SAR models can predict whether adding chelators, radionuclides or chemical groups to a bioactive compound will alter the binding properties and potency of the proposed radiopharmaceutical or cause undesired interactions leading to toxicity or off-target binding. [59,60]

**Behavioral computational modeling**

Absorption, Distribution, Metabolism, Excretion, and Toxicity **(**ADMET) modeling is used to predict the ligand's in vivo behavior based on the pharmacokinetics of similar ligands and modeling interaction of the candidate compound with enzymes involved in toxicity, insolubility, and undesired metabolism [61]. Inputs from SBDD models and SAR, as well as other in silico approaches are used to predict ADMET characteristics. [62] A fundamental step to ensure optimal accuracy of ADMET models is using a chemically diverse data set of ligands to train the model [63].

In addition to proper target specificity and affinity, a translatable radiopharmaceutical should have appropriate pharmacokinetics and stability for optimal performance in vivo. Even slight alterations in radiopharmaceutical structure, such as changes in chelator and linker, can still dramatically affect biodistribution and stability. Hence, using computational ADMET modeling prior to in vivo experiments can help predict serum stability and target organ uptake. [64] Absorption is often not an issue for radiopharmaceuticals as they are generally administrated intravenously. In certain situation such as targeting a receptor in the CNS or an intracellular target ADMET algorithms can be additionally beneficial to model the passage of a radiopharmaceutical through these natural barriers. [7,13,14,65] ADMET models can also help predicting the effect of individual components of the radiopharmaceutical on the overall pharmacokinetics. In this regard, using computer aided drug design (CAAD) models, one can alter the design of a radiopharmaceutical to improve the pharmacokinetics without adversely affecting the binding affinity to the target. [66] This strategy has been successfully implemented for radiolabeled analogs of somatostatin, integrins, bombesin, cholecystokinin and vasoactive intestinal peptide analogues. Lastly, ADMET can help in ensuring that the radiopharmaceutical has minimal toxicity, especially if the ligand is not a naturally occurring molecule or an established pharmaceutical. [67]

**Applications**

To date, molecular docking, SAR and ADMET models are the most commonly used approaches in radiopharmaceutical development. These methods can predict a bioactive compound's interaction with its target after radiolabeling and whether altering radionuclides and chelators affects radiopharmaceutical's performance (figure 1). In this section, we review some of the applications of these methods in radiopharmaceutical design in caner, neurological disorders, and other disease processes.

*Cancer*

Radiopharmaceuticals are extensively used for imaging various nodes in cancer pathways. One common approach is radiolabeling chemotherapy agents to study their pharmacokinetics and their ability to engage the target in the tumor. Molecular docking models have been used to ensure addition of a chelator or a radionuclide does not adversely affect the binding affinity to the target. For instance, a recent study investigated

$^{99m}$technetium ($^{99m}$Tc) labeled Ifosfamide, an alkylating chemotherapy agent, for solid tumor imaging. Molecular docking showed that $^{99m}$Tc labeling did not affect the radiotracer's affinity for the binding site. [38] A similar approach was used with iodine labeled Cladribine, a chlorinated purine analogue, showing that its binding affinity to DNA polymerase was not changed. [68] Another example of using molecular docking is to evaluate effect of changing chelators on binding properties of a commonly used radiopharmaceutical. For instance, Cai et al. showed the in vivo binding affinity of a somatostatin receptor subtype 2 (SSTR2) agonist was not affected by using new generation of chelators for $^{64}$copper ($^{64}$Cu) (Figure 4). [69]

Some studies used a combination of approaches to develop radiopharmaceuticals targeting receptors or enzymes, overexpressed by cancer cells or other cells in tumor microenvironment. An example is development of PET tracers for imaging focal adhesion kinase (FAK), a tyrosine kinase overexpressed in a number of cancers. Inputs from molecular docking and molecular dynamics were utilized to ensure binding of designed inhibitory tracers to FAK as well as the effect of structural modifications, such as altering chain length on inhibitory potency of designed compounds. [70-72] Similar approach was used for targeting gamma glutamyl transferase (GGT) and human farnesyl pyrophosphate synthase (HFPPS). [73,74]

*CNS*

Developing radiopharmaceuticals for imaging different receptors, enzymes and pathologic aggregates in CNS has been of great interest and an area of intense research.

In additional to specificity for the target, a potent radiopharmaceutical for CNS imaging has to cross the blood brain barrier (BBB) and wash out from the non-target tissue fairly rapidly posing an additional challenge. Therefore, the use of in silico methods has gained a lot of attention in CNS radiopharmaceutical development. Novel PET-tracers for cerebral adenosine receptors (AR), phosphodiesterase 2A (PDE2A) and serotonin transporter (SERT) were among the first attempts of using in silico models in CNS radiopharmaceutical design. [41,75,76]

PDE2A is mainly expressed in limbic area and basal ganglia and could be of importance for cognitive function through modulating the signal transduction by a regulating cGMP and cAMP levels. Therefore, development of selective PDE2A PET ligands has been of interest using the in silico models. For this purpose, properties of a large database of successful and failed CNS PET tracers was created and screened to identify favorable ADMET properties for CNS PET imaging. Using these ADMET features, enabled researchers to easily narrow down a library of more than thousand PDE2A inhibitors to less than ten compounds. Next, a SAR model was used to design more potent analogs from the identified lead compound prior to in vivo studies. As predicted by in silico models, the final compound demonstrated optimal performance in vivo. [75] A similar approach was used to design promising radiotracers for PDE4B and nociceptin opioid peptide (NOP) receptor. [77,78]

AR is widely distributed in the brain and involved in different signaling pathways. Therefore, AR PET imaging can be useful for diagnostic and treatment monitoring in a broad range of psychiatric, neurovegetative diseases and Parkinson's disease. A number of $^{11}$carbon ($^{11}$C) labeled tracers and a $^{18}$Fluorine ($^{18}$F) labeled AR tracers have been developed using in silico methods.[79] For instance, [18]F-FESCH which is specific for AR subtype 2A (AR$_{2A}$) receptor has been developed using silico metabolite analysis. This tracer showed high binding affinity for AR$_{2A}$ and favorable pharmacokinetics in preclinical studies. PET imaging of rat brains showed this tracer is accurate for mapping AR$_{2A}$ expression in the brain.[41,80] In silico methods although very powerful, are not always predictive of in vivo behavior of radiopharmaceuticals. Most recently, using QSAR and molecular docking [18]F-TOZ1 was developed based on the structure of tozadenant, an AR$_{2A}$ antagonist with excellent binding affinity. Despite its ability to pass the BBB and high brain uptake, AR$_{2A}$ specific binding was insufficient.[81]

When placing a chelator for radiometal labeling of a lead molecule to image enzymes or cell surface receptors, molecular docking is a useful tool to predict if it affects the binding site to target or adversely affects the binding affinity. Examples are radiotracers developed for imaging cholinesterase, for imaging Alzheimer's disease (AD) or histamine receptor type 1. [37,39]

Pathologic aggregates, such as tau protein, amyloid β (Aβ) and α-synuclein aggregates (α-syn), have been extensively investigated in the pathophysiology of neurodegenerative diseases. [82-84] The complexity of their structure and different molecular isoforms can make tracer design challenging for these targets. [3,85] With the microscopic structure of more aggregates now being discovered, in silico design can accelerate developing high-affinity radiotracers for PET imaging of neurodegenerative diseases. [86,87] In addition, off target binding of these tracers to other targets such as monoamine oxidase B (MAO-B) can be predicted by molecular docking approaches. [88-90] Such information can help optimize binding specificity of next generation of PET radiopharmaceuticals for CNS imaging. Combining different modeling methods to design a merged workflow can improve the ultimate accuracy of modeled data. [83] Further optimization of modeling strategy and variable definition is required to ensure maximal correlation of modeled predictions with in vivo performance. [91] For instance, when different simulation approaches were used to predict binding affinity of styrylbenzoxazole (SB) based tracers for Aβ it highlighted shortcomings of some of the model predictions; molecular dynamics, a complex approach that incorporates particle movements in the model, had higher accuracy than conventional docking models, while quantum mechanical (QM) methods significantly improved prediction accuracy compared to both. Despite higher accuracy, the downside of molecular dynamics and QM is increased complexity and cost of simulations.[83]

*Infection and Inflammation*

Developing non-invasive imaging probes to detect infection has been gaining more attention in the recent years, as structural imaging modalities are insensitive for detection of the site of infection at an early stage. Anti-bacterial antibiotics are attractive compounds for developing radiotracers as they are specific for bacterial targets, allowing for distinction between bacterial infection and inflammation, and have favorable tissue biodistribution for reaching the infected tissue. In addition, their minimal interaction with eukaryotic cells and wash out from normal tissue makes them desirable for infection imaging. [92,93] As the structure of the antibiotics are usually available, chemical similarity search and SAR modeling can be used to screen libraries of antibiotic analogs that have an appropriate binding affinity for imaging purposes. SBDD methods can also be used as the structure of antibiotics' target is commonly known. [92] Ordonez et al. screened commercially available libraries of random radiolabeled small molecules to identify potential substrates involved in bacterial metabolism or interacting with bacteria but with minimal mammalian cell interaction. The identified molecules demonstrated high level of in vitro accumulation in a wide range of bacterial species. Candidate compounds were then labeled with $^{18}$F and successfully identified infection from sterile inflammation in vivo murine models. [94]

Molecular docking has also been used for designing the inflammation PET tracer, $^{68}$Ga-Siglec-9, which targets vascular adhesion protein-1 (VAP-1). Siglec-9 was identified by phage display as a ligand for VAP-1. Molecular docking aided to assess the binding of Siglec-9 to VAP-1 before proceeding with in vivo experiments. [95]

*Other*

SAR models were utilized to predict in vivo performance of three potential radiopharmaceuticals for lung perfusion scan. $^{99m}$Tc-Hexoprenaline, a $β_2$ adrenergic agonist, $^{99m}$Tc-Zolmitriptan, a selective serotonin receptor agonist and $^{131}$I-Dapoxetine, a selective serotonin reuptake inhibitor. Lungs are the reservoirs for Zolmitriptan and Dapoxetine. In the case of Hexoprenaline, the most energetically favored confirmation required addition of $^{99m}$Tc to a moiety essential for appropriate interaction of Hexoprenaline with its target. Hence, poor in vivo binding was predicted for this radiotracer. On the other hand, the position of $^{131}$I in Dapoxetine and $^{99m}$Tc in Zolmitriptan did not affect vital sites for tracer-target interaction, and appropriate target binding was predicted. In vivo experiments in mice confirmed modeled predictions for all three radiopharmaceuticals. [60,96]

**Summary**

In silico approaches are novel tools that guide conventional in vitro and in vivo radiopharmaceutical design experiments and accelerate novel radiopharmaceuticals' bench to bedside translation if used and interpreted appropriately. Various examples of successful incorporation of in silico approaches in radiopharmaceutical design and confirmed validation of modeled data in vivo in a wide range of diseases highlights the additional value of AI integration in this field. However, successful modeling depends on careful inclusion of appropriate variables in the model, the modeling approach, choice of software and availability of accurate structure of the ligands and targets. Thus, developing a systematic workflow could help incorporating computational modeling in routine radiopharmaceutical design process and overcome the current challenges of this valuable technology.

**Clinics Care Points**

- AI can greatly utilize the advances in structural chemistry to accelerate the design of a broad range of radiopharmaceuticals for clinical use.
- In silico approaches can markedly shorten the time frame and reduce the cost associated with radiopharmaceutical development to make them more accessible for non-invasive imaging and therapy in a wide range of targets.
- AI methods have been particularly helpful for CNS radiotracer design, where BBB poses an additional challenge for the radiopharmaceutical to reach its target.
- Currently a well-established systematic approach for incorporating in silico methods in the radiopharmaceutical design workflow is lacking.

- AI is complimentary to the conventional radiopharmaceutical design methods and does not replace them
- In vivo studies and clinical trials are required to confirm utility of in silico designed radiopharmaceuticals.

**Figure legends:**

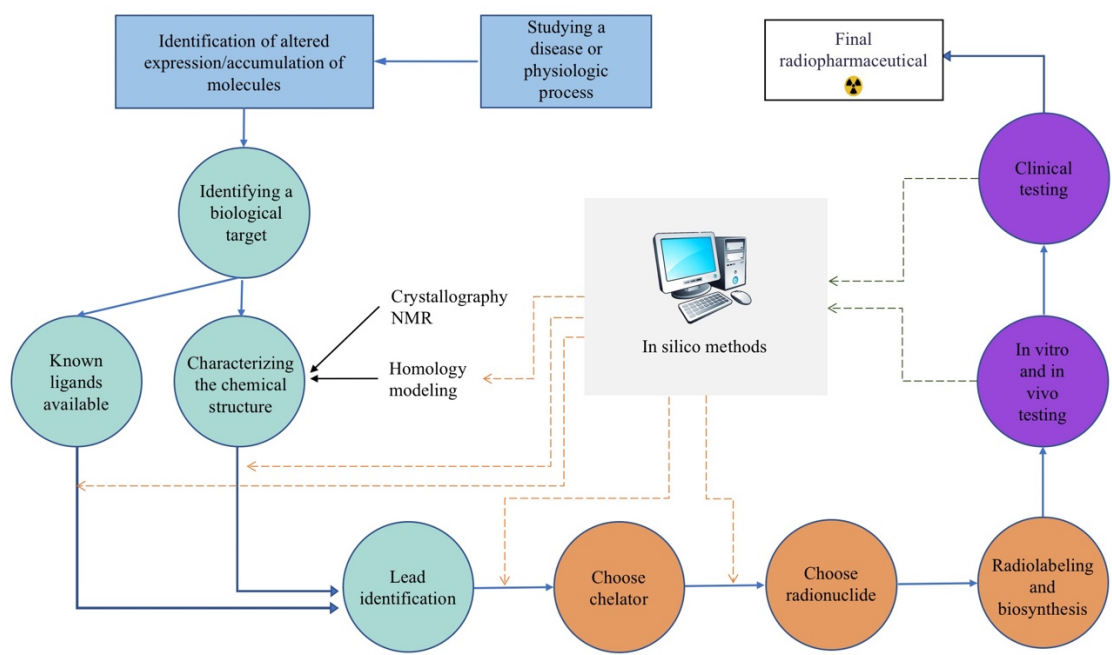

**Figure 1.** Incorporation of in silico drug design methods in routine approach toward developing novel radiopharmaceuticals for unmet clinical needs. *NMR: Nuclear magnetic resonance*

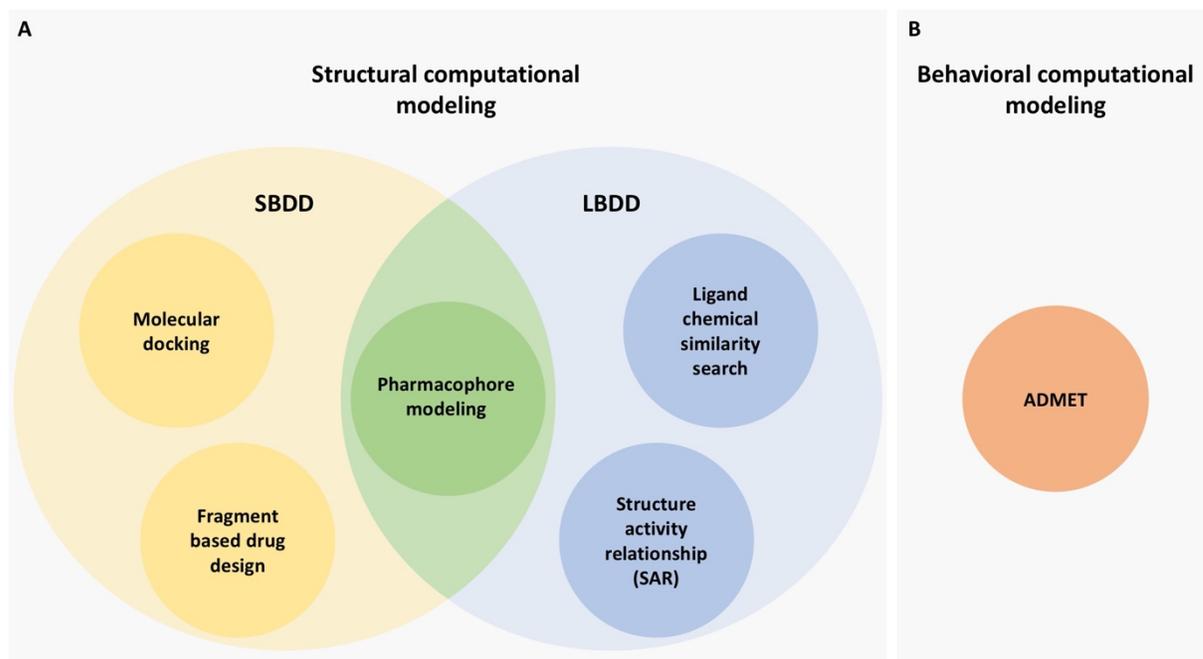

**Figure 2**. Categories of computer aided drug design methods. A) Structural computational modeling, B) Behavioral computational modeling. *SBDD: structure-based drug design, LBDD: ligand-based drug design, ADMET: Absorption, Distribution, Metabolism, Excretion, and Toxicity*

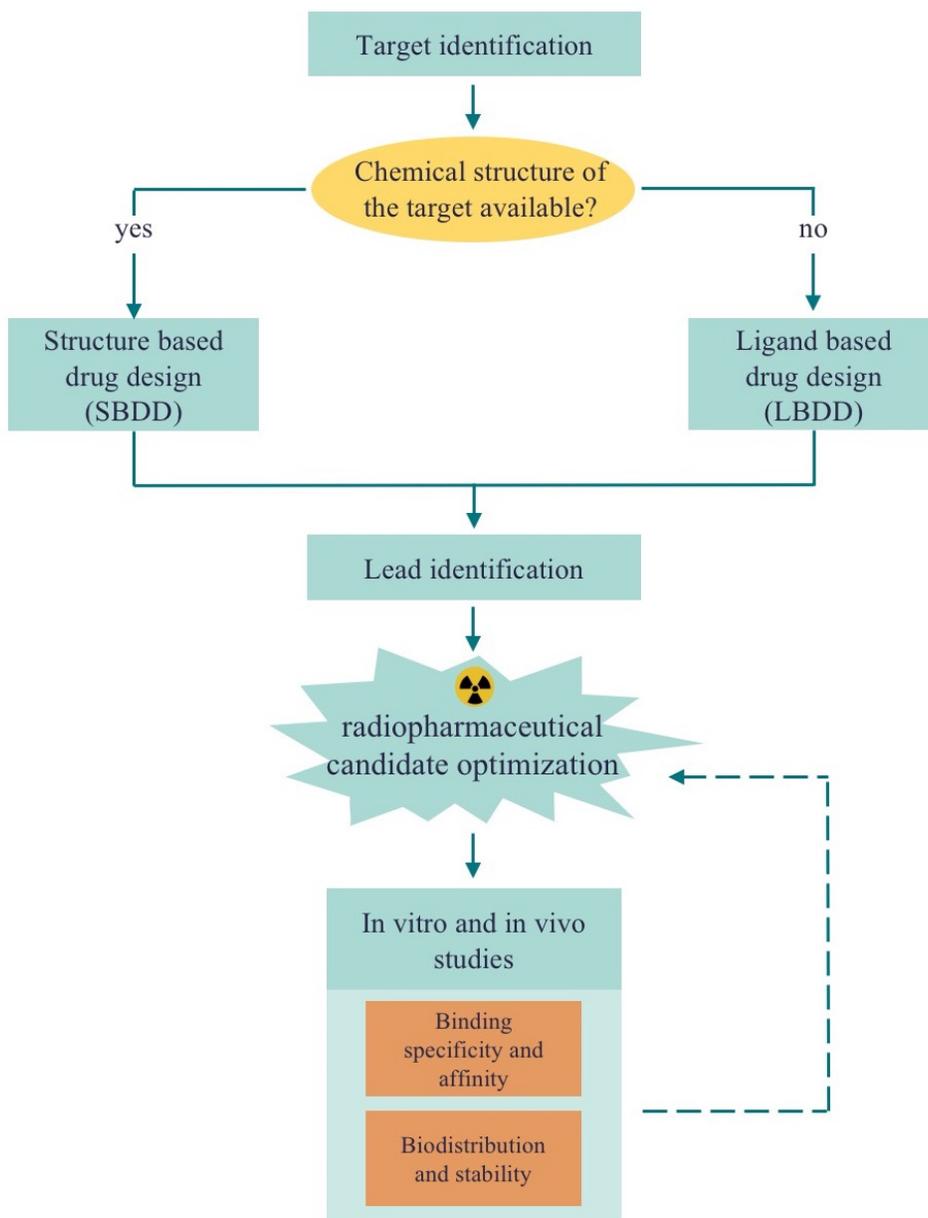

**Figure 3.** Basic workflow of structural computational modeling technics in radiopharmaceutical design and optimization

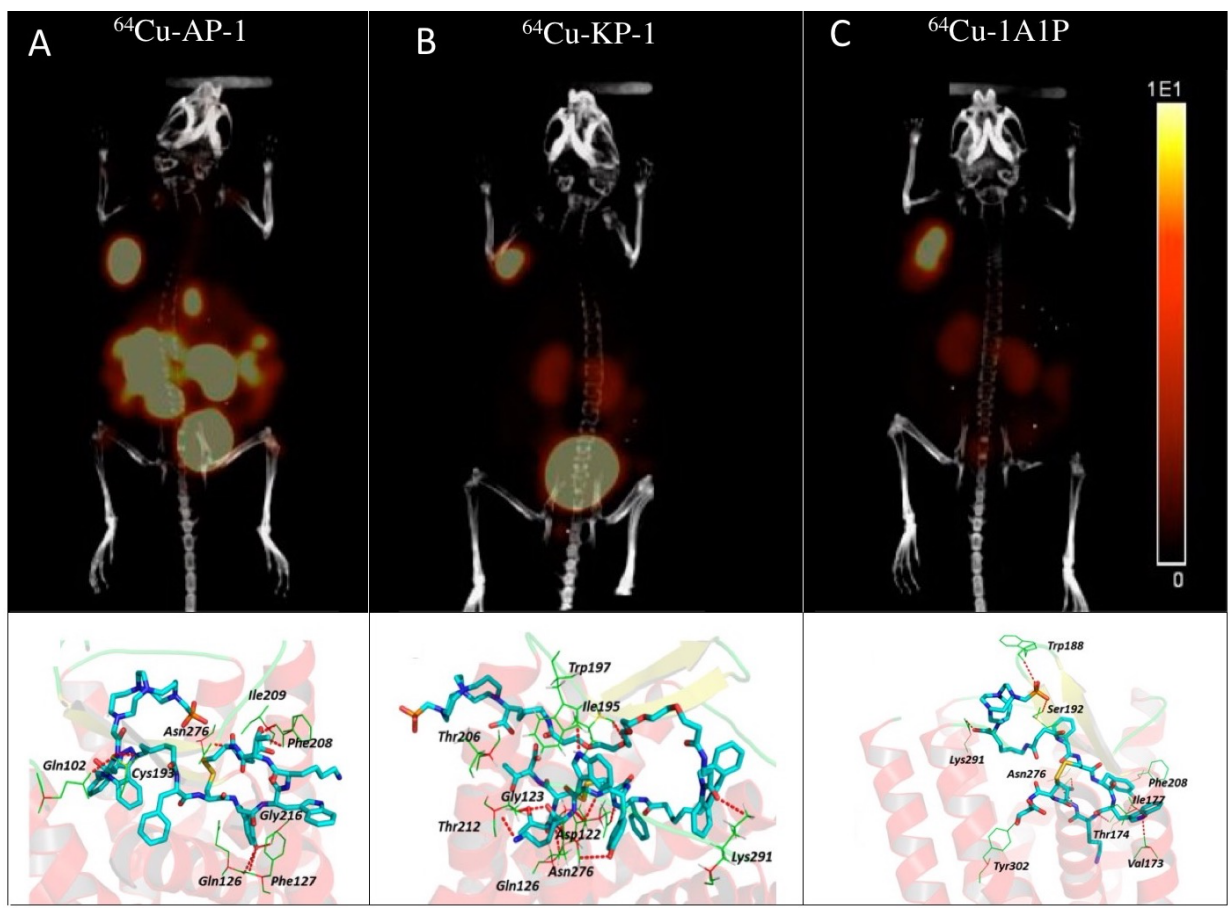

**Figure 4.** Upper panel: Small animal PET/CT images of two probes with a new chelator and labeling approach: A) $^{64}$Cu-AP-1 (SUV = 4.6 ± 0.8) and B) $^{64}$Cu-KP-1 (SUV = 3.2 ± 0.2) compared to an old generation chelator: C) ($^{64}$Cu-1A1P (SUV = 2.8 ± 0.4) shows tumor uptake was only modestly affected by chelator modification, as predicted by molecular docking of each probe in lower panel. However, the biodistribution of the probe was completely altered as shown in PET images. (Content adopted from Reference 69 with publisher's permission. Figures available at link: https://pubs.acs.org/doi/10.1021/jm500416f. Further permission related to the material excerpted should be directed to the publisher.)